\documentclass[
aps,
prl,
reprint,
tightenlines,
superscriptaddress,
nofootinbib,
]{revtex4-2}

\usepackage{booktabs}
\usepackage{makecell}
\usepackage{braket}
\usepackage{amsmath}
\usepackage{eufrak}
\usepackage{graphicx}
\usepackage{subfigure}
\usepackage{grffile}
\usepackage{float}
\usepackage[a4paper, total={6.5in, 9in}]{geometry}
\usepackage[colorlinks,
linkcolor=blue,
anchorcolor=blue,
citecolor=blue
]{hyperref}

\usepackage{xcolor}
\usepackage{comment}
\usepackage{tgtermes}  

\newcommand{\Hz}{\text{Hz}}

\newcommand{\GeV}{\text{GeV}}

\newcommand{\eq}[1]{Eq.~\eqref{#1}}
  \newcommand{\fig}[1]{Fig.~\ref{#1}}
\newcommand{\tab}[1]{Tab.~\ref{#1}}

\newcommand{\REF}[1]{Ref.~\cite{#1}}

\newcommand{\lra}[1]{\left(#1\right)}

\newcommand{\lrd}[1]{\left|#1\right|}

\xdef\figsizeOne{7.5cm}
\xdef\figsizeDouble{8.2cm}

\date{\small\itshape Last update: \today}
\begin{document}
\title{Probing Inflationary Origins of Primordial Black Holes with
LIGO--Virgo--KAGRA O1--O4a data}
\author{Haipeng An}
\email{anhp@mail.tsinghua.edu.cn}
\affiliation{Department of Physics, Tsinghua University, Beijing 100084, China}
\affiliation{Center for High Energy Physics, Tsinghua University, Beijing 100084, China}

\author{Huai-Ke Guo}
\email{guohuaike@ucas.ac.cn}
\affiliation{International Centre for Theoretical Physics Asia-Pacific (ICTP-AP), University of Chinese Academy of Sciences (UCAS), Beijing, China.}
\affiliation{Taiji Laboratory for Gravitational Wave Universe, University of Chinese Academy of Sciences, 100049 Beijing, China}
 
\author{Mai Qiao}
\email{qiaomai@ucas.ac.cn}
\affiliation{International Centre for Theoretical Physics Asia-Pacific (ICTP-AP), University of Chinese Academy of Sciences (UCAS), Beijing, China.}
\affiliation{Taiji Laboratory for Gravitational Wave Universe, University of Chinese Academy of Sciences, 100049 Beijing, China}

\author{Lian-Tao Wang}
\email{liantaow@uchicago.edu}
\affiliation{Department of Physics, The University of Chicago, Chicago, Illinois 60637, USA.}
\affiliation{Enrico Fermi Institute, University of Chicago, Chicago, Illinois 60637, USA}
\affiliation{Kavli Institute for Cosmological Physics, University of Chicago, Chicago, Illinois 60637, USA}

\author{Chen Yang}
\email{yangchen26@ucas.ac.cn}
\affiliation{International Centre for Theoretical Physics Asia-Pacific (ICTP-AP), University of Chinese Academy of Sciences (UCAS), Beijing, China.}
\affiliation{Taiji Laboratory for Gravitational Wave Universe, University of Chinese Academy of Sciences, 100049 Beijing, China}
\affiliation{Department of Physics, Tsinghua University, Beijing 100084, China}

\author{Yue Zhao}
\email{zhaoyue@ust.hk}
\affiliation{Department of Physics and Jockey Club Institute for Advanced Study, The Hong Kong University of Science and Technology, Hong Kong S.A.R., P.R. China}

\begin{abstract}
Large primordial curvature perturbations not only produce primordial black holes (PBHs) but also inevitably source a scalar-induced stochastic gravitational-wave background upon horizon reentry. We analyze the combined LIGO--Virgo--KAGRA O1--O4a data to constrain two representative inflationary mechanisms for generating such perturbations: ultra-slow-roll inflation and an inflationary phase transition. Detecting no evidence for either scenario, we place 
95\% credible upper limits on the curvature-spectrum amplitude across the frequency range accessible to ground-based interferometers. Translated into the PBH context, these limits already exceed conventional constraints, probing abundance fractions far below unity. Our results remain robust even when the PBHs themselves are too rare to be directly detected or have evaporated. This work demonstrates that stochastic gravitational-wave observations offer a powerful and complementary probe of small-scale inflationary physics and PBH formation, with upcoming interferometers promising to extend sensitivity to a wider range of inflationary epochs and PBH masses.
\end{abstract}

\maketitle

{\itshape Introduction}---
Primordial black holes (PBHs) provide a rare observational window onto
primordial density fluctuations at scales far smaller than those accessible
to the cosmic microwave background or large-scale structure \cite{Zeldovich:1967lct,Hawking:1971ei,Carr:1974nx,Carr:1975qj}. In the standard
radiation-dominated cosmology, producing an appreciable PBH abundance requires
a strong enhancement of the primordial curvature power spectrum. When these
enhanced scalar perturbations reenter the horizon, they inevitably source a
stochastic background of gravitational waves (GWs) at second order
\cite{Ananda:2006af,Baumann:2007zm,Kohri:2018awv,Carr:2020gox,Romero-Rodriguez:2021aws}. The GW frequency identifies the
underlying perturbation scale and therefore the characteristic PBH mass,
allowing GW observations to test PBH formation even when the PBHs are too rare
to detect directly or have already evaporated \cite{Saito:2008jc,Yuan:2019udt,Yuan:2021qgz,Jiang:2024aju,Jiang:2026krg}.

There are two widely studied benchmark mechanisms for generating the required enhancement
during inflation. In ultra-slow-roll (USR) inflation, an exceptionally flat
region of the inflaton potential rapidly suppresses the inflaton velocity and
amplifies superhorizon curvature perturbations \cite{Ivanov:1994pa,Kinney:1997ne,Inoue:2001zt,Kinney:2005vj}. In an inflationary phase
transition (InPT), the evolution of the inflaton triggers a first-order
transition in a coupled spectator sector, whose backreaction generates a
localized enhancement of the curvature spectrum
\cite{An:2020fff,An:2022cce,An:2023jxf,An:2023idh}. Although their microscopic dynamics
are distinct, both mechanisms predict correlated PBH and GW signals.

Recent LIGO--Virgo--KAGRA (LVK)
stochastic-background analyses have substantially improved the sensitivity to
cosmological signals
\cite{LIGOScientific:2025kry,Virgo:2025aai,KAGRA:2021kbb,LIGOScientific:2025kbu,LIGOScientific:2026mjf}. Because the PBH abundance depends exponentially on the perturbation amplitude,
even a factor-of-few improvement in that amplitude can translate into
orders-of-magnitude stronger limits on PBH formation.

In this Letter, we use the combined LVK O1--O4a data to test both USR and
InPT. We find no evidence for either signal and derive
95\% credible upper limits on the amplitude as a function of the characteristic frequency of
the scalar perturbations in both scenarios. We further map our results to the constraints on the epoch at which primordial fluctuations were amplified as well as the masses and
abundances of the PBHs they could have produced. We demonstrate that
present GW observations provide a powerful and complementary probe of
PBH-generating dynamics.

{\itshape Early-universe production of GWs and PBHs}---
We consider two representative mechanisms that enhance primordial curvature perturbations on small scales: ultra-slow roll and inflationary phase transition. After horizon reentry, these perturbations source scalar-induced gravitational waves at second order and, if sufficiently large, can collapse to form primordial black holes. In both scenarios, the characteristic frequency of the induced GW spectrum is related to the number of e-folds between the USR or InPT epoch and the end of inflation as
\begin{align}
f_{\rm ref}
    &=
    10^{-9}e^{40-N_e}
    \left(
        \frac{H_{\rm inf}}{10^{14}\,\mathrm{GeV}}
    \right)^{1/2}
    \mathrm{Hz}.
    \label{eq:fref}
\end{align}

1. Ultra-slow roll.---
Ultra-slow-roll inflation occurs when the inflaton traverses an
exceptionally flat region of its potential, such that its evolution is
dominated by Hubble friction,
\begin{equation}
    \ddot{\phi}+3H\dot{\phi}\simeq0 .
\end{equation}
The resulting rapid decrease of the slow-roll parameter amplifies
superhorizon curvature perturbations, for which we adopt the template \cite{Byrnes:2018txb}
\begin{align}
    P_\zeta^{\rm USR}(f)
    &=
    A_{\rm ref}
    \left(\frac{f}{f_{\rm ref}}\right)^4
    \Theta(f_{\rm ref}-f),
    \label{eq:usr-Pzeta}
\end{align}
where $A_{\rm ref}$ is the peak amplitude. The corresponding
present-day scalar-induced GW spectrum is
\begin{equation}
    \Omega_{\rm GW}^{\rm USR}(f)
    =
    \Omega_{\rm R}A_{\rm ref}^{ 2}
    F^{\rm USR}\!\left(\frac{f}{f_{\rm ref}}\right),
    \label{eq:usr-OmegaGW}
\end{equation}
where $\Omega_{\rm R}$ is the present radiation density fraction and the
dimensionless shape function $F^{\rm USR}$ is given in the Supplemental Material.

For the PBH formation, we assume Gaussian
perturbations, Gaussian smoothing, a collapse threshold
$\delta_c=0.45$, a collapse efficiency $\gamma=0.2$, and a monochromatic mass
function centered at the scale of maximum variance \cite{Carr:1975qj,Sasaki:2018dmp, Young:2019yug, Carr:2020gox}.
The characteristic PBH mass and its present dark-matter fraction can be written as
\begin{align}
    M_{\rm PBH}^{\rm USR}
    &=
    3.1\times10^{-17}M_\odot
    \left(\frac{f_{\rm ref}}{\mathrm{Hz}}\right)^{-2},
    \label{eq:usr-MPBH}
    \\
    f_{\rm PBH}^{\rm USR}
    &=
    5.1\times10^{16}
    \left(\frac{f_{\rm ref}}{\mathrm{Hz}}\right)
    \operatorname{erfc}
    \left(\frac{2.1}{\sqrt{A_{\rm ref}}}\right).
    \label{eq:usr-fPBH}
\end{align}

2. Inflationary phase transition.---
An inflationary phase transition can be triggered when the evolution
of the inflaton changes the effective potential of a coupled spectator field
and induces a first-order transition
\cite{An:2020fff,An:2022cce,An:2023jxf}. The backreaction of the transition
on the inflaton can strongly enhance curvature perturbations, which we
parameterize as \cite{An:2023jxf}
\begin{align}
    P_{\zeta}^{\rm InPT}(f)
    &=
    A_{\rm ref}
    \frac{(f/f_{\rm ref})^3}
    {1+(\alpha_1f/f_{\rm ref})^4
      +(\alpha_2f/f_{\rm ref})^9},
    \label{eq:inpt-Pzeta}
\end{align}
where $\alpha_1=0.31$, and $\alpha_2$ depends weakly on the phase-transition rate.
The resulting GW spectrum is
\begin{equation}
    \Omega_{\rm GW}^{\rm InPT}(f)
    =
    \Omega_{\rm R}A_{\rm ref}^{2}
    F^{\rm InPT}\!\left(\frac{f}{f_{\rm ref}}\right),
    \label{eq:inpt-OmegaGW}
\end{equation}
where the dimensionless shape function $F^{\rm InPT}$ is provided in the Supplemental
Material. The corresponding characteristic PBH mass and abundance are
\begin{align}
    M_{\rm PBH}^{\rm InPT}
    &=
    9.7\times10^{-19}M_\odot
    \left(\frac{f_{\rm ref}}{\mathrm{Hz}}\right)^{-2},
    \label{eq:inpt-MPBH}
    \\
    f_{\rm PBH}^{\rm InPT}
    &=
    2.8\times10^{17}
    \left(\frac{f_{\rm ref}}{\mathrm{Hz}}\right)
    \operatorname{erfc}
    \left(\frac{0.26}{\sqrt{A_{\rm ref}}}\right).
    \label{eq:inpt-fPBH}
\end{align}

{\itshape Analysis methods}---Following \REF{Renzini:2023qtj},
we adopt the cross-correlation method to distinguish the correlated SGWB signals from the uncorrelated noise with the point-estimate spectrum $\hat{C}_{IJ}$ of baseline $IJ$ given by~\cite{KAGRA:2021kbb}
\begin{align}
    \hat{C}_{IJ}(f|t)=
    \frac{2\text{Re}[\tilde{d}_I^\star(f|t)\tilde{d}_J(f|t)]}
    {T_{\text{seg}}\gamma_{IJ}(f)S_{IJ}(f)},
\label{eq:cross-correlation_estimator}
\end{align}
where $\tilde{d}_{I(J)}(f)$ is the Fourier transform of the time-domain strain for detector $I(J)$ over a time segment $t$ with length $T_{\text{seg}}$,
$\gamma_{IJ}(f)$ is the normalized overlap reduction function (ORF),
and $S_{IJ}(f)=3\sin\beta_I\sin\beta_J H_0^2/10\pi^2f^3$~\cite{Regimbau:2012ir},
where $H_0$ is the Hubble constant and $\beta_{I(J)}$ is the opening angle between the two arms of detector $I(J)$.
Specifically,
the opening angles are $\pi/2$ and $\pi/3$ for L-shaped detectors such as LVK and V-shaped detectors such as ET,
respectively~\cite{Regimbau:2012ir}.
In the weak-signal limit,
the variance of the point-estimate spectrum given by \eq{eq:cross-correlation_estimator} can be approximately expressed as
\begin{align}
    \sigma_{IJ}^2(f|t)
    \approx
    \frac{1}{2T_{\text{seg}}\Delta f}
    \frac{P_I(f|t)P_J(f|t)}
    {\gamma_{IJ}^2(f)S_{IJ}^2(f)}
    ,
    \label{eq:cvar}
\end{align}
where $P_{I(J)}$ is the power spectral density~(PSD) of the detector $I(J)$,
and $\Delta f$ is the frequency resolution.
We use the inverse-variance weighting to combine the point-estimate spectrum for each time segment and baseline as follows to suppress fluctuations of the estimator spectrum,
\begin{align}
\begin{split}
    \hat{C}_{IJ}(f)
    =&
    \frac
    {
    \sum_{t}
    \hat{C}_{IJ}(f|t)
    \sigma_{IJ}^{-2}(f|t)
    }
    {
    \sum_{t}
    \sigma_{IJ}^{-2}(f|t)
    }
    ,
\end{split}
\end{align}
where $\sigma_{IJ}^{2}(f)=1/\sum_{t}\sigma_{IJ}^{-2}(f|t)$ is the combined variance spectrum.
In this work,
we use the publicly available package \texttt{pygwb}~\cite{Renzini:2023qtj} to calculate the cross-correlation spectra and their variances.

To infer SGWB signals from the measured cross-correlation spectrum between two detectors,
we perform Bayesian inference with the signal-parameter-dependent part of the likelihood function given by~\cite{Allen:1997ad,Mandic:2012pj}
\begin{align}
\begin{split}
     \ln p
=
-\sum_{IJ,\, k}
\frac{
(\hat{C}_{IJ}
    \lra{f_k}
    -
    \lambda_{IJ}
    \Omega_{\text{GW}}(f_k|\boldsymbol{\Theta})
    )^2
    }
    {2\sigma_{IJ}^2(f_k)}
,
\end{split}
\end{align}
where $\boldsymbol{\Theta}$ are the parameters of signal models and $\lambda_{IJ}$ is a factor that compensates for the systematic error from the calibration process~\cite{Sun:2020wke},
which is marginalized according to the methods described in \REF{Renzini:2023qtj,Whelan:2012ur}.

We take $\boldsymbol{\Theta}=(f_{\text{ref}},\,A_{\text{ref}})$ as free parameters for the USR and InPT models in Bayesian inference.
Apart from the SGWB generated during early cosmology, we also include the astrophysical SGWB, which is expected to be dominated by unresolved CBCs. We model the GW energy-density spectrum from these unresolved sources as a PL with a fixed spectral index of $2/3$ and a reference amplitude $\Omega_{\text{ref}}$ at $25~\Hz$~\cite{Virgo:2025aai,KAGRA:2021kbb}.

In \tab{tab:priors},
we summarize the priors of the parameters we choose during the Bayesian inference process,
where, notably,
the upper bound of $A_{\text{ref}}$ ensures that the total fractional energy density does not exceed unity.

\begin{table}[H]\centering
    \begin{tabular}{@{}lc@{}}
\hline\hline
    \textbf{Parameter} & \textbf{Prior Distribution} \\
    \hline
    $\Omega_{\text{ref}}$ & $\text{Log10Uniform}(10^{-13},\,10^{-6})$ \\
    $f_{\text{ref}}$ (InPT/USR) & $\text{Log10Uniform}(0.1,\,10^{4})~\Hz$ \\
    $A_{\text{ref}}$ (USR) & $\text{Log10Uniform}(10^{-6},\,80)$ \\
    $A_{\text{ref}}$ (InPT) & $\text{Log10Uniform}(10^{-8},\,10)$ \\
    \hline\hline
    \end{tabular}
    \caption{
Prior distributions used for Bayesian inference in this work.
}
    \label{tab:priors}
\end{table}

To project the sensitivity of next-generation ground-based GW detectors such as ET~\cite{Branchesi:2023mws,Punturo:2010zz,Hild:2010id,ET:2025xjr} and CE~\cite{Evans:2021gyd,Reitze:2019iox} to the SGWBs induced by USR or InPT,
we define the threshold of exclusion by setting the signal-to-noise ratio~(SNR) as $\rho=2$.
For a given SGWB model and interferometer-based detector pair, the SNR is given by
\begin{align}
\rho=
\sqrt{
        2T
        \int_{f_{\min}}^{f_{\max}}
        d f 
        S_{IJ}^2(f)
        \frac{\gamma_{IJ}^2(f)\Omega^2_{\text{GW}}(f)}
        {
P_{n,\,I}(f)P_{n,\,J}(f)
        }
},
\label{eq:snr}
\end{align}
where $P_{n,\,I(J)}(f)$ is the noise PSD for detector $I$($J$)~\cite{LIGO:T1500293},
and $T$ is the total observation time.
For future space-based detectors~\cite{Hu:2017mde,Ruan:2018tsw,LISACosmologyWorkingGroup:2022jok,LISA:2017pwj,TianQin:2015yph,TianQin:2020hid},
we adopt the absolute SNR for the A channel defined in \REF{Liang:2026wwz} to estimate the sensitivity of SGWB detection.
The frequency-integration ranges adopted for the SNR calculation are summarized in the Supplemental Material.

\begin{figure}[ht!]
\centering
    \includegraphics[width=\figsizeDouble]{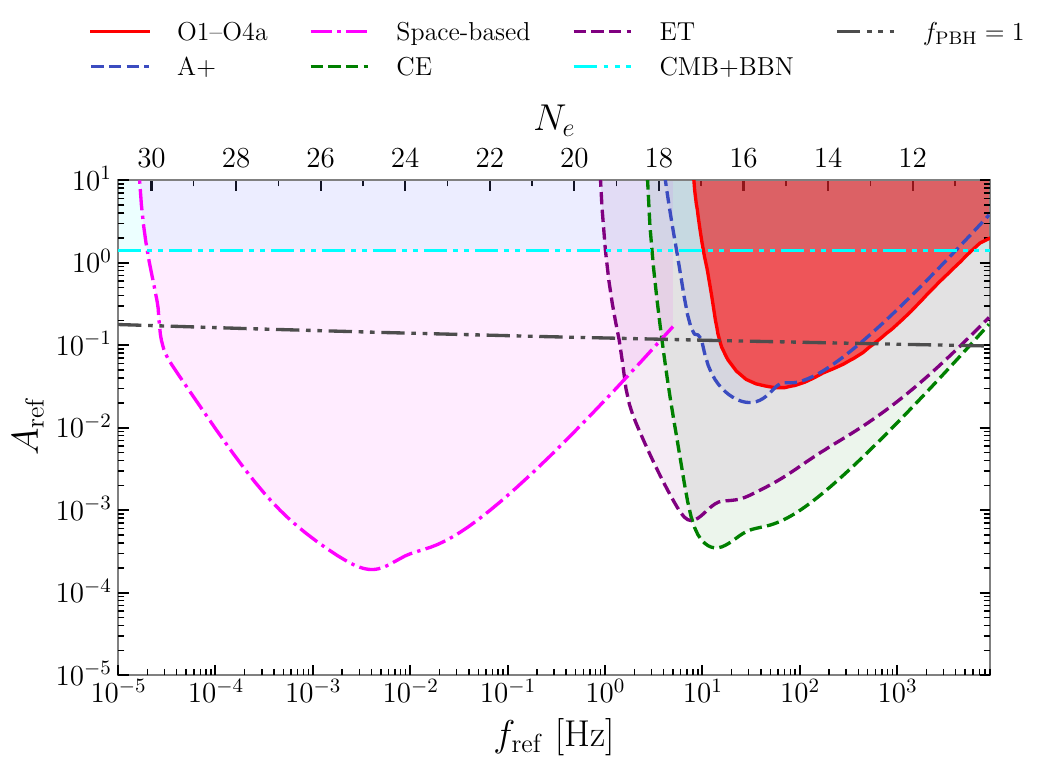}
    \includegraphics[width=\figsizeDouble]{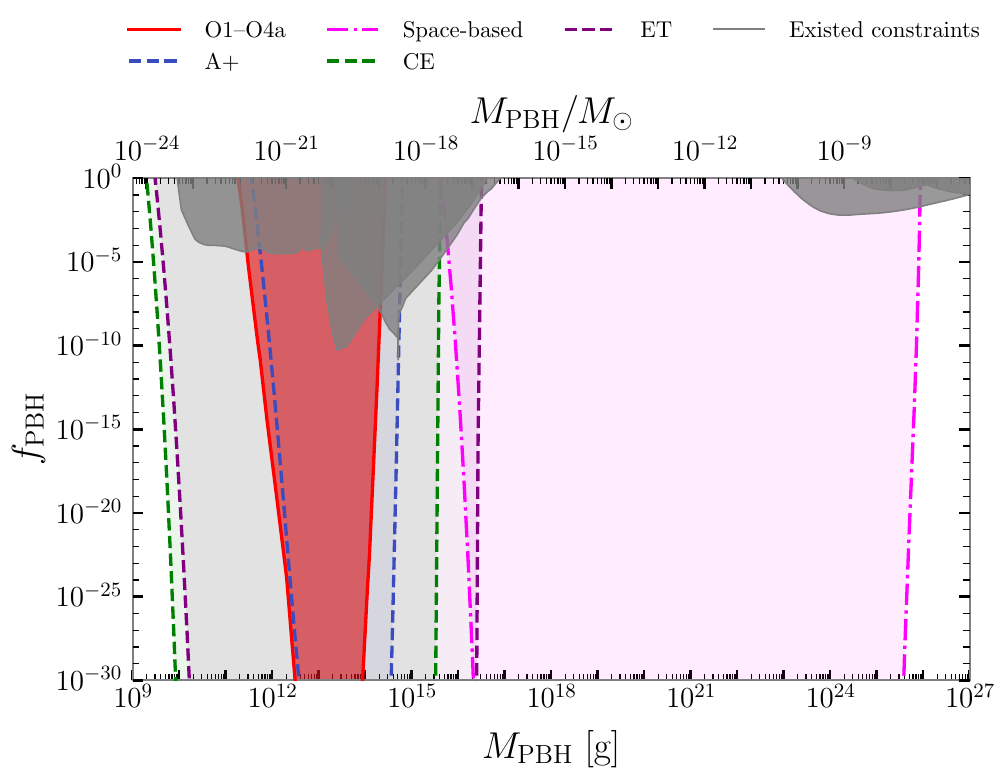}
    \caption{
    \textit{Constraints on ultra-slow-roll inflation.}
\textit{Upper:} The $95\%$ credible upper limit from the combined LVK
O1--O4a data on the curvature-power-spectrum amplitude $A_{\rm ref}$ as a
function of the reference frequency $f_{\rm ref}$. Also shown are the joint
CMB+BBN bound~\cite{Yeh:2022heq}, the contour corresponding to $f_{\rm PBH}=1$, and the
projected one-year sensitivities (${\rm SNR}=2$) of A+ and the
future GW experiments~\cite{LIGO:T1500293,Liang:2026wwz}.
The upper axis translates $f_{\rm ref}$ into the
number of e-folds $N_e$ between the end of the USR phase and the end of
inflation, assuming $H_{\rm inf}=10^{14}\,\GeV$.
\textit{Lower:} The corresponding GW bounds translated into the
characteristic PBH mass $M_{\rm PBH}$ and dark-matter fraction
$f_{\rm PBH}$ using Eqs.~\eqref{eq:usr-MPBH} and
\eqref{eq:usr-fPBH}. Existing constraints from BBN~\cite{Carr:2009jm},
the CMB~\cite{Acharya:2020jbv,Chluba:2020oip},
Hawking evaporation~\cite{Carr:2009jm,Carr:2016hva,Boudaud:2018hqb},
and microlensing~\cite{Smyth:2019whb,Griest:2013esa,Griest:2013aaa,EROS-2:2006ryy} are included for comparison. 
    }
    \label{fig:sen_usr}
\end{figure}

\begin{figure}[ht!]
\centering
    \includegraphics[width=\figsizeDouble]{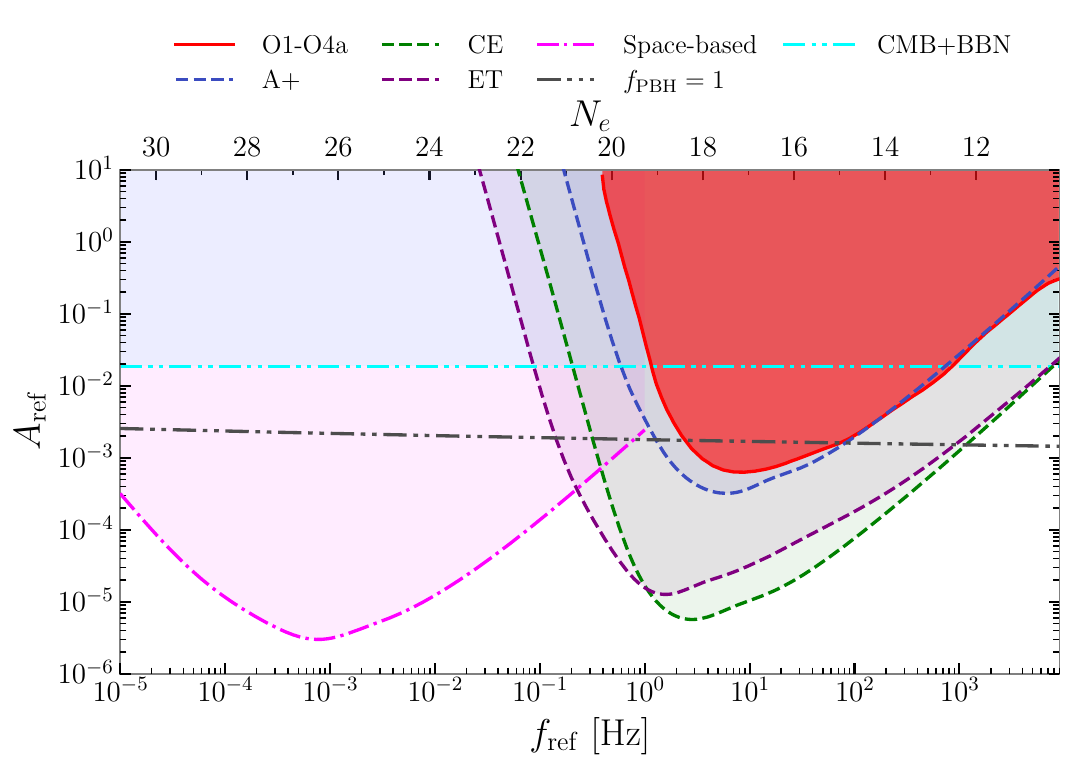}
    \includegraphics[width=\figsizeDouble]{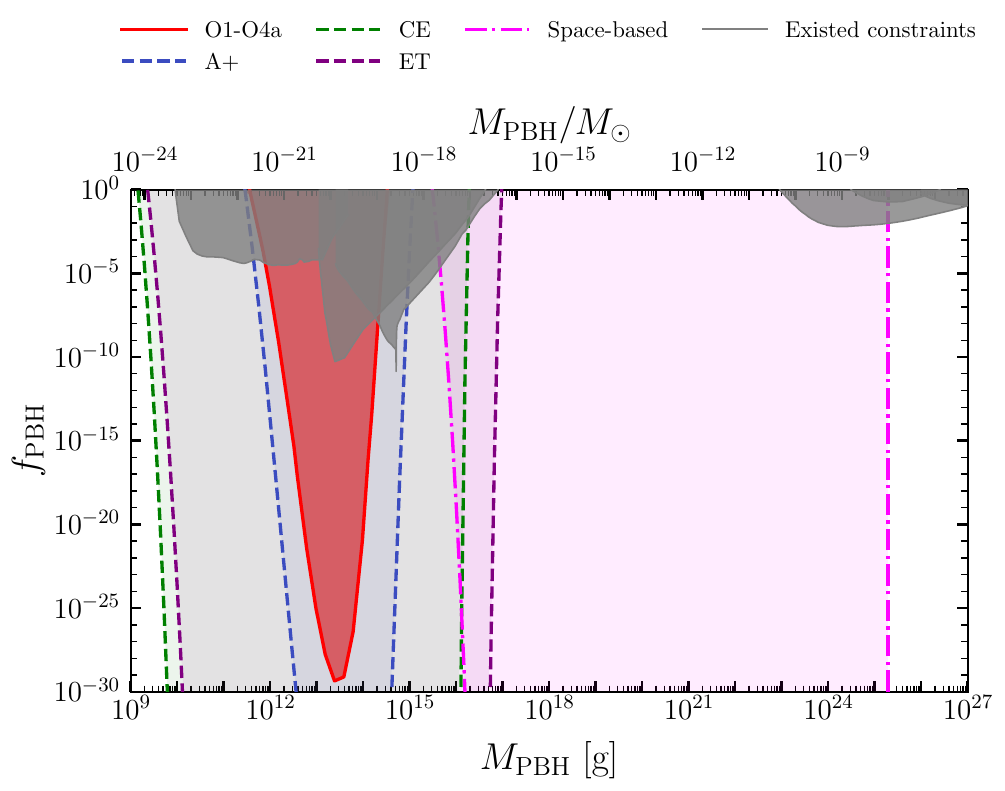}
    \caption{
\textit{Constraints on Inflationary Phase Transition.}
   Same as Fig.~\ref{fig:sen_usr}, but for the InPT scenario. We show the
constraints on $A_{\rm ref}$ and $f_{\rm ref}$ (\textit{upper}) and their
translation into the PBH parameter space (\textit{lower}). In the upper
panel, $f_{\rm ref}$ is mapped to the number of e-folds $N_e$ between the
phase transition and the end of inflation. 
    }
    \label{fig:sen_inpt}
\end{figure}

{\itshape Data}---In this work,
we use the data from the O1--O4a runs of LVK detectors.
More explicitly, O1--O4a data contain data from the LIGO-Livingston~(L) and LIGO-Hanford detectors~(H)~\cite{LIGOScientific:2016gtq,LIGO:2021ppb,Capote:2024rmo,LIGO:2024kkz} and O3 data also include data from the Virgo~(V) detector~\cite{Virgo:2022ysc,VIRGO:2014yos}.
For O1--O3 data,
we use the publicly available cross-correlation spectra of the HL baseline for the O1--O3 run and of HV and LV baselines for the O3 run~\cite{LIGOScientific:2021b,KAGRA:2021kbb}.
We treat the cross-correlation spectra from the same two detectors but different observing runs as different baselines.
For the newly published O4a data,
we use the strain data from the \texttt{GDS-CALIB\_STRAIN\_CLEAN\_AR} channel of the H and L detectors.
The total duration of the synchronized HL data before applying data-quality vetoes is about $126.57$ days~\cite{Virgo:2025aai}.
We apply the ``\emph{Category 1}'' veto to remove data collected during periods when the detectors were not operating in nominal conditions~\cite{LIGOScientific:2017tza,LIGO:2024kkz}.
The remaining synchronized data are further divided into 192-s segments with $50\%$ overlap between consecutive segments~\cite{Renzini:2023qtj}.
Both time-domain and frequency-domain cuts are applied to the research data to ensure the strain data are stationary and not contaminated by loud instrumental artifacts.
Among them,
the time-domain cuts consist of gating and delta-sigma cut (DSC)~\cite{KAGRA:2021kbb,Virgo:2025aai}.
We use the \texttt{pygwb}~\cite{Renzini:2023qtj} package to apply the gating procedures aiming to remove glitches from the time-domain data without losing too much data by multiplying the time series by an inverse Planck-taper window~\cite{Virgo:2025aai}.
The DSC aims to remove the segments whose variances defined in \eq{eq:cvar} fluctuate more than $20~\%$ relative to those of the nearby segments~\cite{KAGRA:2021kbb}.
The DSC is performed four times, with $\alpha=-5$, $0$, $3$, and $5$, using the variance weight $(f/f_{\text{ref}})^{-2\alpha}$ in each case~\cite{KAGRA:2021kbb,Virgo:2025aai}.
For the frequency-domain cut, 
we notch the frequency bins containing identified instrumental artifacts including calibration lines~\cite{Karki:2016pht}, 
quadruple suspension violin modes~\cite{LIGOScientific:2014pky},
powerline harmonics,
and other known instrumental artifacts for each segment of O4a data~\cite{Virgo:2025aai}.

{\itshape Results}---
We find no evidence for a scalar-induced gravitational-wave background from
either USR or InPT in the combined LVK O1--O4a data. We therefore derive
$95\%$ credible upper limits on the curvature-spectrum amplitude
$A_{\rm ref}$ at fixed reference frequency $f_{\rm ref}$ for each scenario,
marginalizing over a stochastic background from unresolved compact-binary
coalescences. Figures~\ref{fig:sen_usr} and
\ref{fig:sen_inpt} show the resulting constraints in the
$(f_{\rm ref},A_{\rm ref})$ plane and their corresponding interpretation in
the $(M_{\rm PBH},f_{\rm PBH})$ plane.

In \tab{tab:o1o4a-stronger-than-fpbh-one}, we summarize the parameter ranges in which the LVK O1--O4a constraints are stronger than the PBH abundance bound, $f_{\rm PBH}\leq 1$.

\providecommand{\USRMinFref}{\ensuremath{56.2}}
\providecommand{\USRMinNe}{\ensuremath{15.2}}
\providecommand{\USRMinMPBH}{\ensuremath{9.8\times10^{-21}}}
\providecommand{\USRMinMPBHGrams}{\ensuremath{1.9\times10^{13}}}
\providecommand{\USRMinMPBHTenElevenGrams}{\ensuremath{194.9}}
\providecommand{\USRMinAref}{\ensuremath{3.1\times10^{-2}}}
\providecommand{\USRMinFPBH}{\ensuremath{2.3\times10^{-46}}}

\providecommand{\InPTMinFref}{\ensuremath{8.9}}
\providecommand{\InPTMinNe}{\ensuremath{17.1}}
\providecommand{\InPTMinMPBH}{\ensuremath{1.2\times10^{-20}}}
\providecommand{\InPTMinMPBHGrams}{\ensuremath{2.4\times10^{13}}}
\providecommand{\InPTMinMPBHTenElevenGrams}{\ensuremath{242.8}}
\providecommand{\InPTMinAref}{\ensuremath{6.3\times10^{-4}}}
\providecommand{\InPTMinFPBH}{\ensuremath{4.5\times10^{-30}}}

\providecommand{\USRStrongFrefMin}{\ensuremath{15.8}}
\providecommand{\USRStrongFrefMax}{\ensuremath{562.3}}
\providecommand{\USRStrongNeMin}{\ensuremath{12.9}}
\providecommand{\USRStrongNeMax}{\ensuremath{16.5}}
\providecommand{\USRStrongMPBHMin}{\ensuremath{9.8\times10^{-23}}}
\providecommand{\USRStrongMPBHMax}{\ensuremath{1.2\times10^{-19}}}
\providecommand{\USRStrongMPBHGramsMin}{\ensuremath{1.9\times10^{11}}}
\providecommand{\USRStrongMPBHGramsMax}{\ensuremath{2.5\times10^{14}}}
\providecommand{\USRStrongMPBHTenElevenGramsMin}{\ensuremath{1.9}}
\providecommand{\USRStrongMPBHTenElevenGramsMax}{\ensuremath{2.5\times10^{3}}}

\providecommand{\InPTStrongFrefMin}{\ensuremath{2.8}}
\providecommand{\InPTStrongFrefMax}{\ensuremath{70.8}}
\providecommand{\InPTStrongNeMin}{\ensuremath{15.0}}
\providecommand{\InPTStrongNeMax}{\ensuremath{18.2}}
\providecommand{\InPTStrongMPBHMin}{\ensuremath{1.9\times10^{-22}}}
\providecommand{\InPTStrongMPBHMax}{\ensuremath{1.2\times10^{-19}}}
\providecommand{\InPTStrongMPBHGramsMin}{\ensuremath{3.8\times10^{11}}}
\providecommand{\InPTStrongMPBHGramsMax}{\ensuremath{2.5\times10^{14}}}
\providecommand{\InPTStrongMPBHTenElevenGramsMin}{\ensuremath{3.8}}
\providecommand{\InPTStrongMPBHTenElevenGramsMax}{\ensuremath{2.5\times10^{3}}}

\providecommand{\CESNRFmin}{\ensuremath{5}}
\providecommand{\CESNRFmax}{\ensuremath{5\times10^{3}}}
\providecommand{\ETSNRFmin}{\ensuremath{1}}
\providecommand{\ETSNRFmax}{\ensuremath{10^{4}}}
\providecommand{\APlusSNRFmin}{\ensuremath{5}}
\providecommand{\APlusSNRFmax}{\ensuremath{5\times10^{3}}}
\providecommand{\SpaceBasedSNRFmin}{\ensuremath{3\times10^{-5}}}
\providecommand{\SpaceBasedSNRFmax}{\ensuremath{0.5}}
\providecommand{\SpaceBasedSNRObservationDays}{\ensuremath{365}}
\providecommand{\SpaceBasedSNRObservationSeconds}{\ensuremath{31536000}}
\providecommand{\CESNRFrequencyRange}{\ensuremath{\CESNRFmin\mbox{--}\CESNRFmax\,\mathrm{Hz}}}
\providecommand{\ETSNRFrequencyRange}{\ensuremath{\ETSNRFmin\mbox{--}\ETSNRFmax\,\mathrm{Hz}}}
\providecommand{\APlusSNRFrequencyRange}{\ensuremath{\APlusSNRFmin\mbox{--}\APlusSNRFmax\,\mathrm{Hz}}}
\providecommand{\SpaceBasedSNRFrequencyRange}{\ensuremath{\SpaceBasedSNRFmin\mbox{--}\SpaceBasedSNRFmax\,\mathrm{Hz}}}
\providecommand{\TaijiSNRFmin}{\SpaceBasedSNRFmin}
\providecommand{\TaijiSNRFmax}{\SpaceBasedSNRFmax}
\providecommand{\TaijiSNRObservationDays}{\SpaceBasedSNRObservationDays}
\providecommand{\TaijiSNRObservationSeconds}{\SpaceBasedSNRObservationSeconds}
\providecommand{\TaijiSNRFrequencyRange}{\SpaceBasedSNRFrequencyRange}

\providecommand{\InPTLogBF}{\ensuremath{-0.315}}
\providecommand{\InPTLogBFError}{\ensuremath{0.003}}

\providecommand{\CBCInPTLogBF}{\ensuremath{-0.780}}
\providecommand{\CBCInPTLogBFError}{\ensuremath{0.004}}

\providecommand{\USRLogBF}{\ensuremath{-0.192}}
\providecommand{\USRLogBFError}{\ensuremath{0.003}}

\providecommand{\CBCUSRLogBF}{\ensuremath{-0.652}}
\providecommand{\CBCUSRLogBFError}{\ensuremath{0.004}}
 
\begin{table}[htbp]
    \centering
    \begin{tabular}{lccc}
        \hline\hline
        \textbf{Model}
        & $f_{\rm ref}\,[\mathrm{Hz}]$
        & $N_e$
        & $M_{\rm PBH}\,[10^{11}\,\mathrm{g}]$ \\
        \hline
        USR
        & $\bigl(\USRStrongFrefMin,\ \USRStrongFrefMax\bigr)$
        & $\bigl(\USRStrongNeMin,\ \USRStrongNeMax\bigr)$
        & $\bigl(\USRStrongMPBHTenElevenGramsMin,\ \USRStrongMPBHTenElevenGramsMax\bigr)$ \\
        InPT
        & $\bigl(\InPTStrongFrefMin,\ \InPTStrongFrefMax\bigr)$
        & $\bigl(\InPTStrongNeMin,\ \InPTStrongNeMax\bigr)$
        & $\bigl(\InPTStrongMPBHTenElevenGramsMin,\ \InPTStrongMPBHTenElevenGramsMax\bigr)$ \\
        \hline\hline
    \end{tabular}
    \caption{Ranges along the $95\%$ credible LVK O1--O4a upper-limit curves, for which the LVK constraint is stronger than $f_{\rm PBH}= 1$. The corresponding values of $N_e$ are obtained assuming $H_{\rm inf}=10^{14}\,\mathrm{GeV}$.
    }
    \label{tab:o1o4a-stronger-than-fpbh-one}
\end{table}

Here we see the particular power of GW observations as a
probe of PBH formation. A
factor-of-few improvement in $A_{\rm ref}$ can therefore become an
improvement of many orders of magnitude in $f_{\rm PBH}$.

Furthermore, because the scalar-induced GW background is produced at horizon reentry, it remains observable irrespective of PBH survival to the present epoch. Given that the relevant PBH masses are below the threshold for survival to today, our limits should be viewed as constraints on PBH formation. The gravitational-wave signal retains this formation information even after PBH evaporation, allowing LVK observations to access early-Universe parameter space that is beyond the reach of present-day PBH searches.

The projected sensitivities of future detectors extend this
reach to smaller scalar amplitudes and a broader range of inflationary
e-folds; because of the exponential amplitude--abundance relation, their
improvement in PBH parameter space will be substantially larger than the
corresponding improvement in $A_{\rm ref}$.

{\itshape Conclusion and outlook.---}
In this study, we use the combined LVK O1--O4a data to examine two widely studied
mechanisms for producing large primordial scalar perturbations:
ultra-slow-roll inflation and an inflationary phase transition. Although their
microscopic dynamics are different, both mechanisms can generate PBHs together
with a correlated scalar-induced gravitational-wave background. The absence of
such a signal therefore places constraints, for both scenarios, on the inflationary dynamics responsible for PBH
production. Remarkably, the current O1--O4a data are already sufficiently
powerful to surpass existing PBH limits over part of the relevant parameter
space and to probe the PBH abundance fraction far below unity.

The gravitational-wave spectrum retains simultaneous information about when the
USR or InPT occurred during inflation and which PBH mass scale it would have
produced. Moreover, because PBH formation is exponentially sensitive to the
amplitude of the scalar perturbations, even a modest improvement in the
gravitational-wave limit can examine vast regions of PBH parameter space. This
makes stochastic-background searches particularly powerful for masses at
which conventional PBH observations are weak, and even for PBHs that have
already evaporated. In this sense, the gravitational-wave background provides a channel to otherwise inaccessible small-scale structure in the
primordial Universe.

Future interferometers, spanning a broader frequency range and
reaching substantially lower background amplitudes, will extend this program
across new inflationary epochs and PBH masses. Gravitational-wave observations can become a central probe
of inflation, primordial black holes, and the thermal history of the early
Universe.

{\itshape Acknowledgment}---H.A. is supported by the NSFC under Grant Nos. 12475107 and 12525506, and the National Key R\&D Program of China under Grant Nos. 2023YFA1607104 and 2021YFC2203100.
This work is partly supported by the National Natural Science Foundation of China (NSFC) under Grant No. 12347103, 12547104 and W2611007. L.T.W. is supported by the
Department of Energy under Grant No. DE-SC0013642.

This research has made use of data or software obtained from the Gravitational Wave Open Science Center (gwosc.org), a service of the LIGO Scientific Collaboration, the Virgo Collaboration, and KAGRA. This material is based upon work supported by NSF's LIGO Laboratory which is a major facility fully funded by the National Science Foundation, as well as the Science and Technology Facilities Council (STFC) of the United Kingdom, the Max-Planck-Society (MPS), and the State of Niedersachsen/Germany for support of the construction of Advanced LIGO and construction and operation of the GEO600 detector. Additional support for Advanced LIGO was provided by the Australian Research Council. Virgo is funded, through the European Gravitational Observatory (EGO), by the French Centre National de Recherche Scientifique (CNRS), the Italian Istituto Nazionale di Fisica Nucleare (INFN) and the Dutch Nikhef, with contributions by institutions from Belgium, Germany, Greece, Hungary, Ireland, Japan, Monaco, Poland, Portugal, Spain. KAGRA is supported by Ministry of Education, Culture, Sports, Science and Technology (MEXT), Japan Society for the Promotion of Science (JSPS) in Japan; National Research Foundation (NRF) and Ministry of Science and ICT (MSIT) in Korea; Academia Sinica (AS) and National Science and Technology Council (NSTC) in Taiwan.

\bibliographystyle{apsrev4-1}
\bibliography{bib}

\appendix

\clearpage

\section{Details of GW and PBH production from curvature perturbations}

\subsection{USR}

During single-field inflation, if the inflaton potential becomes very flat, the power spectrum will grow rapidly. This is called ultra-slow-roll inflation \cite{Ivanov:1994pa,Kinney:1997ne,Inoue:2001zt,Kinney:2005vj}. We have
\begin{equation}
    \ddot{\phi}+3H\dot{\phi}\simeq0\ ,
\end{equation}
then the slow-roll parameters satisfy
\begin{align}
    \eta&\simeq-6\ ,\\
    \epsilon&\propto e^{-6N_{\rm USR}}\ ,
\end{align}
where $N_{\rm USR}$ is the e-folding number of USR. During USR, the scalar tilt $n_s-1\simeq4$, so we can parametrize the power spectrum as a power law \cite{Byrnes:2018txb},
\begin{equation}
    P_\zeta^{\rm USR}=A_{\rm ref}(f/f_{\rm ref})^4\Theta(f_{\rm ref}-f)\ ,
\end{equation}
where $A_{\rm ref}$ is the peak amplitude of the USR power spectrum,
which is determined by the duration of USR, and
\begin{equation}
    f_{\text{ref}}
    =
    10^{-9}
    e^{40-N_e}
    \left(
        \frac{H_{\text{inf}}}{10^{14}~\GeV}
    \right)^{1/2}\ {\rm Hz}\ ,
\end{equation}
where $N_e$ is the number of e-folds between the end of USR and the end of inflation, and $H_{\rm inf}$ is the Hubble parameter of inflation. 

The large curvature perturbations will induce SIGWs as well as PBHs when they reenter the horizon. The energy density spectrum of SIGWs is given by
\begin{equation}
    \begin{split}
    \Omega_{\rm GW}(f)
    =&
    \Omega_{\rm R}
    \frac{f^3}{6}
    \int_{-1}^1
    d\mu
    \int_0^\infty
    df_1
    K\lra{f,\,f_1,\,f_2}
    \\
    &P_\zeta(f_1)
    P_\zeta(f_2)
    (1-\mu^2)^2
    \lra{\frac{f_1}{f_2}}^3,
    \label{eq:f}
\end{split}
\end{equation}
where $f_2=\sqrt{f^2+f_1^2-2ff_1\mu}$. The kernel function $K$ is given by \cite{Kohri:2018awv}
\begin{align}
\begin{split}
    K
    =&
    \frac{f_3^4}{2}
    \lra{\frac{3}{4f_1^3f_2^3}}^2
    \left(
    \pi^2
    f_3^4
    \Theta(f_1+f_2-\sqrt{3}f)
    \right.
    \\
    &+
    \left.
    \lra{
    4f_1f_2
    +
    f_3^4
    \ln
    \lrd{
        \frac{3f^2-(f_1-f_2)^2}{3f^2-(f_1+f_2)^2}
    }
    }^2
    \right)
    ,
\end{split}
\end{align}
where $\Theta$ is the Heaviside step function and $f_3=f_1^2+f_2^2-3f^2$. 

\begin{figure}[ht!]
\centering
    \includegraphics[width=\figsizeDouble]{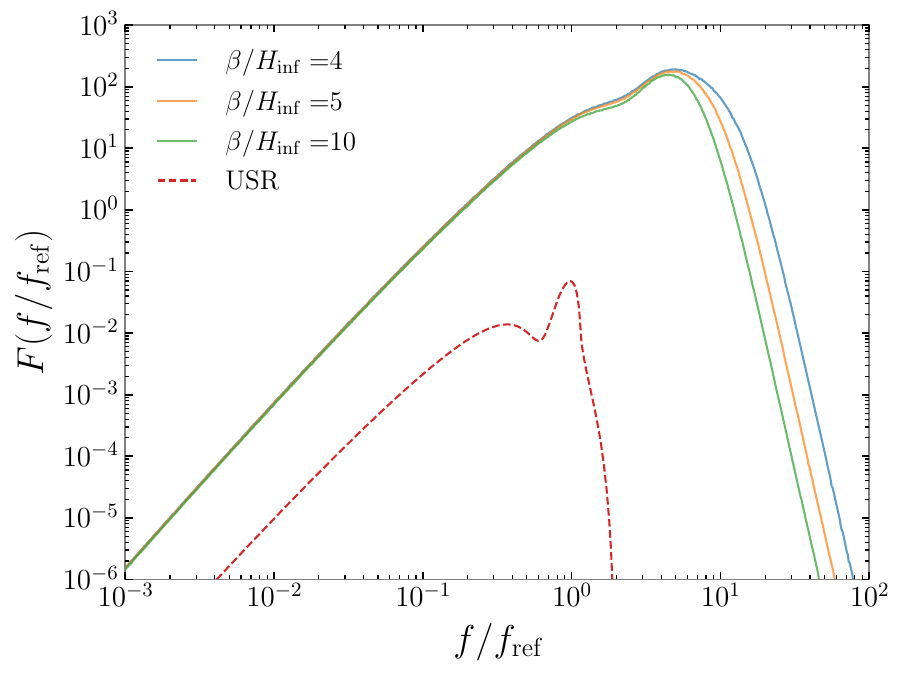}
\caption{
Here we show $F^{\rm USR}$ and $F^{\rm InPT}$ as functions of $f/f_{\text{ref}}$. We note that, for InPT, $F^{\rm InPT}$ has a very mild dependence on $\beta/H_{\text{inf}}$.
    }
    \label{fig:f_func}
\end{figure}

Therefore, the GW spectrum induced by USR is given by
\begin{equation}
    \Omega_{\rm GW}^{\rm USR}(f)=\Omega_{\rm R}A_{\rm ref}^{2} F^{\rm USR}(f/f_{\rm ref})\ ,
\end{equation}
where $F^{\rm USR}$ is shown in Fig.~\ref{fig:f_func}.

When curvature perturbations reenter the horizon, if the energy contrast $\delta$ exceeds a critical value, the perturbation will collapse to form a PBH. Assuming Gaussianity, the variance of energy contrast for a given comoving horizon scale $R=1/(aH)$ is given by
\begin{equation}
    \sigma_R^2=\int_0^\infty \frac{df}{f}\frac{16}{81}(2\pi f R)^4 P_\zeta(f)W_R(f)^2\ ,
\end{equation}
where $W_R$ is the window function. We adopt a Gaussian window function
\begin{equation}
    W_R(f)=\exp\left(-\frac{1}{2}(2\pi f R)^2\right)\ .
\end{equation}
The PBH mass can be estimated by the horizon mass,
\begin{equation}
    M_{\rm PBH}=\gamma \frac{aR}{2G}\ ,
\end{equation}
where $G$ is the Newton constant and $\gamma$ can be chosen as $0.2$ \cite{Carr:1975qj}. The PBH energy fraction at formation is given by the tail of a Gaussian distribution \cite{Carr:2020gox},
\begin{equation}
    \Omega_{\rm PBH}|_{\rm formation}={\rm erfc}\left(\frac{\delta_c}{\sqrt{2}\sigma_R}\right)\ ,
\end{equation}
where ${\rm erfc}$ is the complementary error function and $\delta_c$ is the critical value of energy contrast for PBH formation. Here we choose $\delta_c=0.45$ \cite{Young:2019yug}. 

In principle, a primordial power spectrum can produce PBHs at any mass scale.
In this work,
for simplicity,
we consider a monochromatic PBH mass function,
with the mass determined by the scale at which the variance is maximal.
Therefore, the comoving horizon scale and variance are given by
\begin{align}
    R_{\rm max}&=\frac{1.7}{2\pi f_{\rm ref}}\ ,\\
    \sigma^2_{\rm max}&=0.023A_{\rm ref}\ .
\end{align}
Then we have
\begin{align}
    M_{\rm PBH}^{\rm USR}&=3.1\times10^{-17}M_\odot\cdot\left(\frac{f_{\rm ref}}{\rm Hz}\right)^{-2}\ ,\\
    f_{\rm PBH}^{\rm USR}&\equiv\left.\frac{\Omega_{\rm PBH}}{\Omega_{\rm DM}}\right|_{\rm formation} \nonumber\\
    &=5.1\times10^{16}\cdot\left(\frac{f_{\rm ref}}{\rm Hz}\right){\rm erfc}\left(\frac{2.1}{\sqrt{A_{\rm ref}}}\right)\ .
\end{align}

\subsection{InPT}
During slow-roll inflation, the large excursion of the inflaton field, $\phi$, often approaching the Planck scale, can induce substantial changes in the properties of a spectator field $\sigma$ coupled to $\phi$, such as its effective mass squared and couplings. As a result, the evolution of the inflaton can trigger phase transitions. 

We consider the simple model
\begin{equation}
    \mathcal{L}
    = -\frac{1}{2}\left(\partial_\mu \phi\right)^2
      -\frac{1}{2}\left(\partial_\mu \sigma\right)^2
      - c_m \phi^2 \sigma^2
      - V_0(\phi)
      - V_1(\sigma)\,,
\end{equation}
where inflation is governed by $V_0(\phi)$, and the phase transition can be either first- or second-order, depending on the shape of $V_1(\sigma)$. This model has been studied in detail in \cite{An:2020fff,An:2022cce,An:2023idh,An:2023jxf}.

The backreaction on the inflaton can generate large curvature perturbations. For a first-order phase transition, the power spectrum is given by~\cite{An:2023jxf}
\begin{equation}
    P_{\zeta}^{\rm InPT}(f)=A_{\rm ref}\frac{(f/f_{\rm ref})^3}{1+(\alpha_1 f/f_{\rm ref})^4+(\alpha_2 f/f_{\rm ref})^9}\ ,
    \label{eq:Pinpt}
\end{equation}
where 
\begin{equation}
\begin{aligned}
    A_{\text{ref}}
    &=
    \frac{24}{\epsilon}
    \left(
        \frac{M_{\text{pl}}}{\phi_0}
    \right)^2
    \left(
        \frac{H_{\text{inf}}}{\beta}
    \right)^3
    \left(
        \frac{L}{\rho_{\text{inf}}}
    \right)^2
    \,,
    \\
    f_{\text{ref}}
    &=
    10^{-9}
    e^{40-N_e}
    \left(
        \frac{H_{\text{inf}}}{10^{14}~\GeV}
    \right)^{1/2}
    ~\Hz
    \,,
\end{aligned}
\label{eq:aref_fref}
\end{equation}
respectively. Here, $\beta$ is the time derivative of the bounce action, $L$ is the latent energy density, $\epsilon$ is the slow-roll parameter, $\rho_{\rm inf}$ is the total energy density of the Universe during inflation, and $N_e$ denotes the number of e-folds between the phase transition and the end of inflation. The numerical parameter $\alpha_1$ is fixed to 0.31~\cite{An:2023jxf},
and $\alpha_2$ depends slightly on $\beta/H_{\text{inf}}$ as shown in \tab{tab:alpha1_alpha2}.

\begin{table}[htbp]
    \centering
    \begin{tabular}{ccccc} 
        \hline\hline
        {$\beta/H_{\text{inf}}$} & {$4$} & {$5$} & {$10$} & {$20$} \\
        \midrule
        $\alpha_{2}$ & 0.14 & 0.17 & 0.20 & 0.20 \\
        \hline\hline
    \end{tabular}
    \caption{$\alpha_{2}$ in \eq{eq:Pinpt} for selected values of $\beta/H_{\text{inf}}$~\cite{An:2023jxf}.}
    \label{tab:alpha1_alpha2}
\end{table}

Similarly, using \eq{eq:f}, we have
\begin{equation}
    \Omega_{\rm GW}^{\rm InPT}(f)=\Omega_{\rm R}A_{\rm ref}^2 F^{\rm InPT}(f/f_{\rm ref})\ .
\end{equation}
The function $F^{\rm InPT}$ is shown in~\fig{fig:f_func}.

For InPT, the maximum variance and corresponding comoving horizon scale are given by
\begin{align}
    R_{\rm max}&=\frac{0.3}{2\pi f_{\rm ref}}\ ,\\
    \sigma^2_{\rm max}&=1.5A_{\rm ref}\ .
\end{align}
Therefore,
\begin{align}
    M_{\rm PBH}^{\rm InPT}=&9.7\times10^{-19}M_\odot\cdot\left(\frac{f_{\rm ref}}{\rm Hz}\right)^{-2}\ ,\\
    f_{\rm PBH}^{\rm InPT}
    =&2.8\times10^{17}\cdot\left(\frac{f_{\rm ref}}{\rm Hz}\right){\rm erfc}\left(\frac{0.26}{\sqrt{A_{\rm ref}}}\right)\ .
\end{align}

\section{Additional information on the data analysis}

\begin{figure*}[ht!]
    \centering
    \includegraphics[width=\figsizeOne]{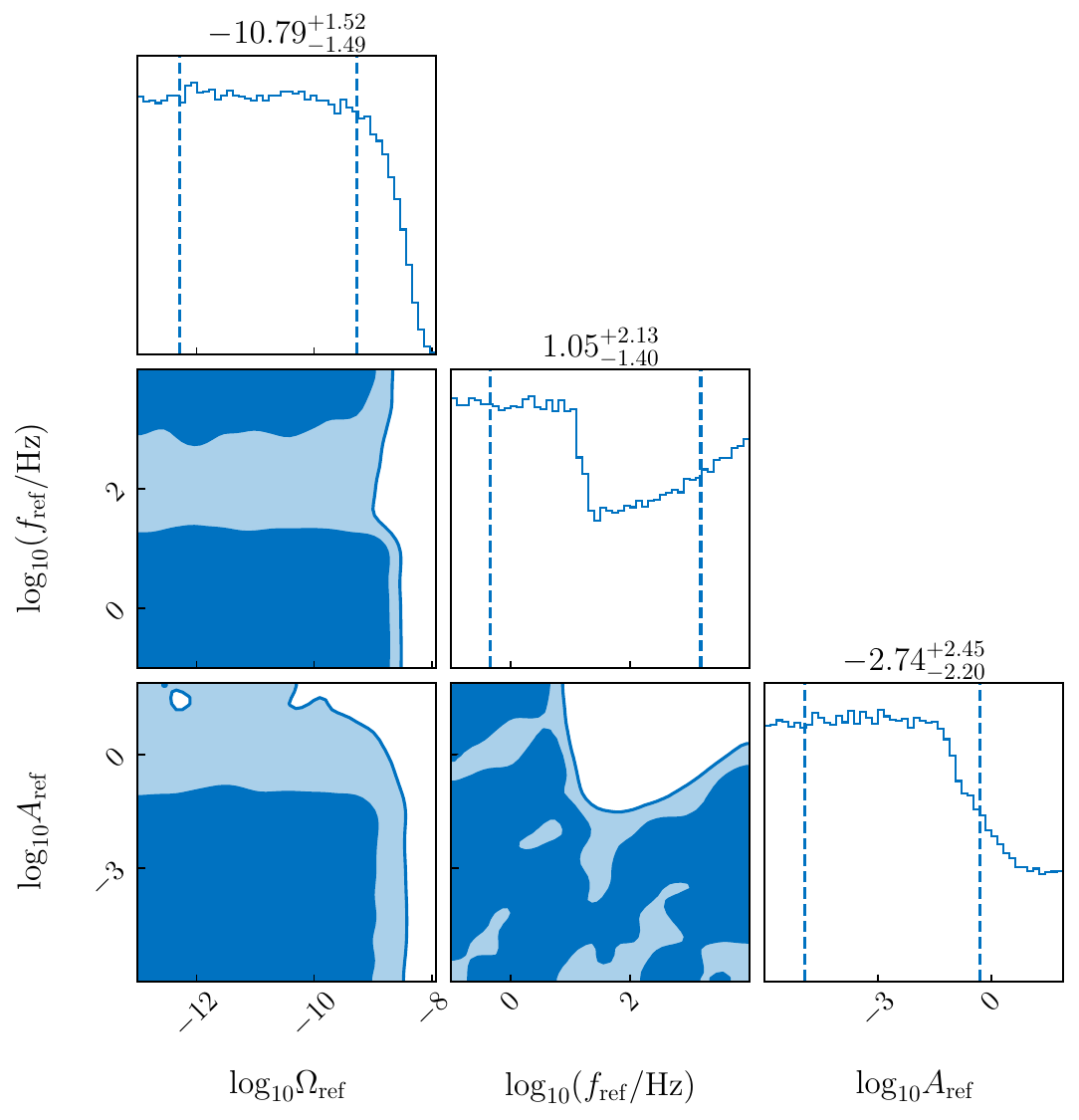}
    \includegraphics[width=\figsizeOne]{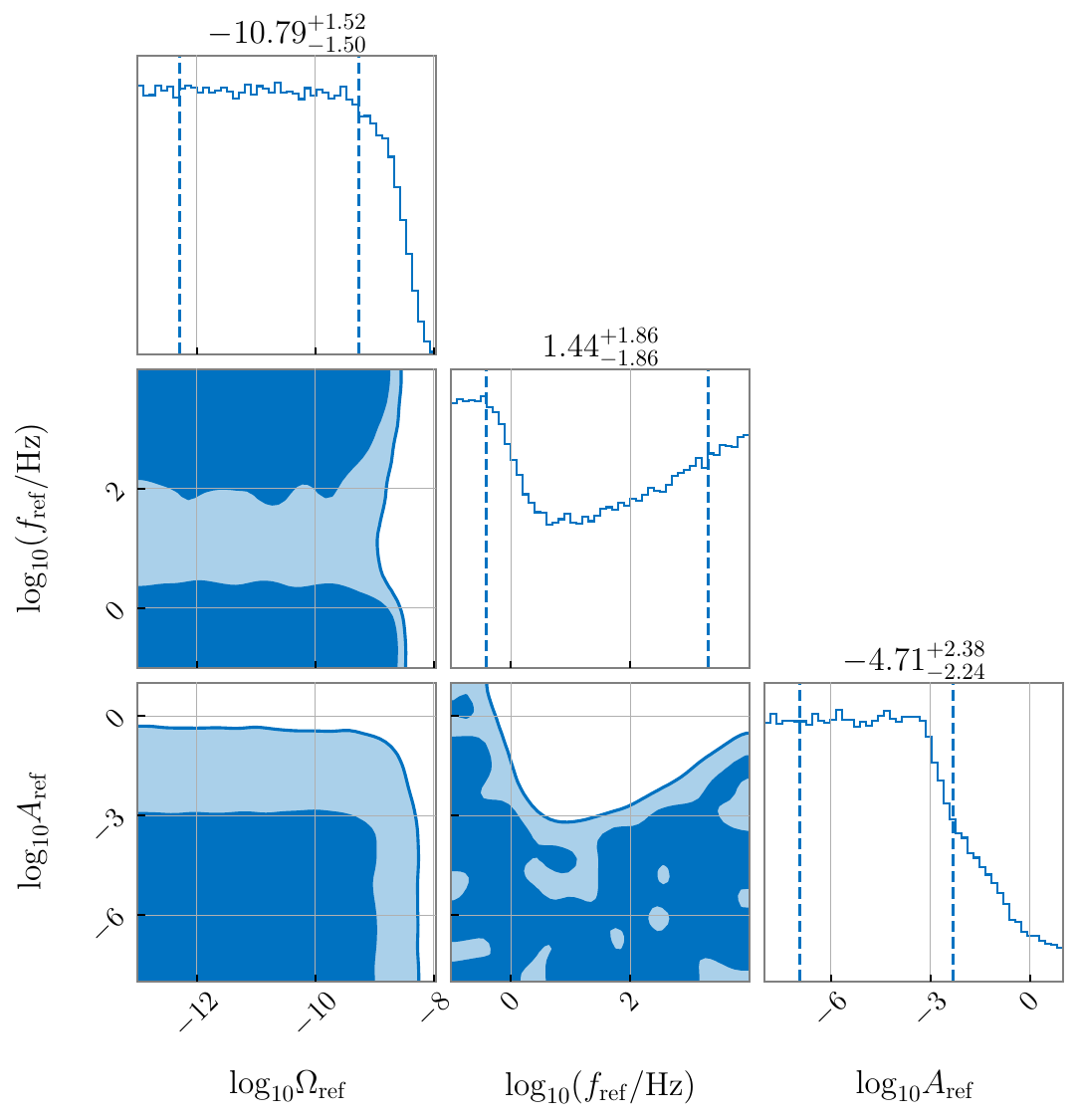}
    \caption{
    \textit{Left:}
Marginalized posterior probability distribution of the combined model of CBCs and USR using the LVK O1--O4a data for parameters $\Omega_\text{ref}$, $f_\text{ref}$ and $A_\text{ref}$ in $\log_{10}$ space.
The blue-dashed vertical lines stand for the $16\%$ and $84\%$ percentiles, respectively.
The dark-blue (light-blue) contours enclose the 68\% (95\%) credible regions.
\textit{Right:}
Same as the left panel,
but for the combined CBC+InPT model with $\beta/H_{\text{inf}}=5$.
}
    \label{fig:corner_sim_cbc_usr}
\end{figure*}

In this section, we provide additional information on the Bayesian analysis and the projected SNR contours presented in this work.

In \fig{fig:corner_sim_cbc_usr},
we show the marginalized posterior distributions for parameters $\Omega_\text{ref}$, $f_\text{ref}$ and $A_\text{ref}$ in $\log_{10}$ space of the combined CBC+USR and CBC+InPT models,
respectively, inferred from the LVK O1--O4a data.

\begin{table}[htbp]
    \centering
    \begin{tabular}{lc}
        \hline\hline
        Model & $\ln \mathcal{B}_{\rm S/N}$ \\
        \hline
        InPT & $\InPTLogBF \pm \InPTLogBFError$ \\
        CBC+InPT & $\CBCInPTLogBF \pm \CBCInPTLogBFError$ \\
        USR & $\USRLogBF \pm \USRLogBFError$ \\
        CBC+USR & $\CBCUSRLogBF \pm \CBCUSRLogBFError$ \\
        \hline\hline
    \end{tabular}
    \caption{Natural logarithms of the Bayes factors, together with their uncertainties, for the four signal models relative to the noise-only model, as inferred from the O1--O4a data.}
    \label{tab:o1o4a-log-bayes-factors}
\end{table}

In \tab{tab:o1o4a-log-bayes-factors},
we summarize the log Bayes factors for four signal models relative to the noise-only model, inferred from the LVK O1--O4a data.
The four signal models are InPT, USR, CBC+InPT, and CBC+USR.
The negative log Bayes factors indicate that the data do not favor any of these signal models over the noise-only model.

\begin{table}[htbp]
    \centering
    \begin{tabular}{lcccc}
        \hline\hline
        & A+ & CE & ET & Space-based \\
        \hline
        $f_{\min}\,[\mathrm{Hz}]$
        & \APlusSNRFmin
        & \CESNRFmin
        & \ETSNRFmin
        & \SpaceBasedSNRFmin \\
        $f_{\max}\,[\mathrm{Hz}]$
        & \APlusSNRFmax
        & \CESNRFmax
        & \ETSNRFmax
        & \SpaceBasedSNRFmax \\
        \hline\hline
    \end{tabular}
    \caption{Frequency integration ranges used to calculate the SNR
    sensitivity curves in this work.}
    \label{tab:snr-frequency-integration-ranges}
\end{table}

In \tab{tab:snr-frequency-integration-ranges},
we summarize the frequency ranges used to calculate the projected SNR contours for future GW observatories.

\end{document}